\RequirePackage{lineno}
\documentclass[letterpaper,twocolumn,prl,aps,superscriptaddress,amsmath,amssymb,floatfix]{revtex4-2}
\usepackage{mathptmx}
\usepackage[utf8]{inputenc}
\usepackage{color}
\usepackage{amsmath}
\usepackage{amssymb}
\usepackage{graphicx}
\usepackage{esint}
\usepackage[unicode=true,
 bookmarks=true,bookmarksnumbered=false,bookmarksopen=false,
 breaklinks=false,pdfborder={0 0 1},backref=false,colorlinks=true]
 {hyperref}
\hypersetup{
 linkcolor=magenta,urlcolor=blue,citecolor=blue,pdfstartview={FitH},hyperfootnotes=false}

\makeatletter

\DeclareFontEncoding{LGR}{}{}

\ProvideTextCommand{\~}{LGR}[1]{\char126#1}

\newcommand{\lyxmathsym}[1]{\ifmmode\begingroup\def\b@ld{bold}
  \text{\ifx\math@version\b@ld\bfseries\fi#1}\endgroup\else#1\fi}

\usepackage{textcomp}
\usepackage{epstopdf}

\usepackage{amsfonts}

\pdfpageheight\paperheight
\pdfpagewidth\paperwidth

\@ifundefined{textcolor}{}{%
 \definecolor{BLACK}{gray}{0}
 \definecolor{WHITE}{gray}{1}
 \definecolor{RED}{rgb}{1,0,0}
 \definecolor{GREEN}{rgb}{0,1,0}
 \definecolor{BLUE}{rgb}{0,0,1}
 \definecolor{CYAN}{cmyk}{1,0,0,0}
 \definecolor{MAGENTA}{cmyk}{0,1,0,0}
 \definecolor{YELLOW}{cmyk}{0,0,1,0}
}

\usepackage{xcolor}\usepackage{soul}
\definecolor{blue}{rgb}{0,0,1}
\definecolor{red}{rgb}{1,0,0}
\definecolor{green}{rgb}{0,1,0}

\makeatother

\begin{document}

%\setpagewiselinenumbers
%\modulolinenumbers[5]
%\linenumbers

% \title{Suppressing the decoherence of an optically trapped single atom: \\ Realization of 16.6-s coherence time with a single cesium atom in an optical dipole trap}
\title{Extending the coherence time limit of a single-alkali-atom qubit by suppressing phonon-jumping-induced decoherence}
\affiliation{State Key Laboratory of Quantum Optics and Quantum Optics Devices, and
Institute of Opto-Electronics, Shanxi University, Taiyuan 030006,
China}
\affiliation{Collaborative Innovation Center of Extreme Optics, Shanxi University,
Taiyuan 030006, China}

\author{Zhuangzhuang Tian}
\thanks{These authors contributed equally to this work.}
\author{Haobo Chang}
\thanks{These authors contributed equally to this work.}
\author{Xin Lv}
\thanks{These authors contributed equally to this work.}
\author{Mengna Yang}
\author{Zhihui Wang}
\author{Pengfei Yang}
\author{Pengfei Zhang}
\author{Gang Li}
\email{gangli@sxu.edu.cn}
\author{Tiancai Zhang}
\email{tczhang@sxu.edu.cn}
\affiliation{State Key Laboratory of Quantum Optics and Quantum Optics Devices, and
Institute of Opto-Electronics, Shanxi University, Taiyuan 030006,
China}
\affiliation{Collaborative Innovation Center of Extreme Optics, Shanxi University,
Taiyuan 030006, China}

%\date{\today}

\begin{abstract}
In the fields of quantum metrology and quantum information processing with the system of optically trapped single neutral atoms, the coherence time of qubit encoded in the electronic states is regarded as one of the most important parameters. Longer coherence time is always pursued for higher precision of measurement and quantum manipulation. The coherence time is usually assumed to be merely determined by relative stability of the energy between the electronic states, and the analysis of the decoherence was conducted by treating the atom motion classically. We proposed a complete description of the decoherence of a qubit encoded in two ground electronic states of an optically trapped alkali atom by adopting a full description of the atomic wavefunction. The motional state, i.e., the phonon state, is taken into account. In addition to decoherence due to the variance of differential light shift (DLS), a new decoherence mechanism, phonon-jumping-induced decoherence (PJID), was discovered and verified experimentally. The coherence time of a single-cesium-atom qubit can be extended to $T_2\approx 20$ s by suppressing both the variances of DLS and PJID by trapping the atom in a blue-detuned bottle beam trap (BBT) and preparing the atom in its three-dimensional motional ground states. The coherence time is the longest for a qubit encoded in an optically trapped single alkali atom. Our work provides a deep understanding of the decoherence mechanism for single atom qubits and thus provides a new way to extend the coherence time limit. The method can be applied for other atoms and molecules, opening up new prospects for high-precision control the quantum states of optically trapped atoms or molecules.
\end{abstract}

\maketitle

\section{Introduction}
The systems of optically trapped neutral single atoms are important platforms for quantum metrology \cite{kaufman_3s_clock,kaufman_30s_clock,Kauffman_spin_squeeze,Antoine_spin_squeeze}, quantum measurement \cite{Tian_Xia_EDM_Ra,Tian_Xia_EDM_Yb}, quantum computation \cite{Saffman_algorithms,lukin_algorithms}, and quantum simulations \cite{2D_Antiferromagnets,Continuous_symmetry_breaking,Spin_liquids}.
Usually, ground-state Zeeman sublevels, such as clock states, are adopted to encode the information as qubits for applications. The coherence time ($T_2$) of the qubit is thus one of the key factors for the high performance in these applications, and a long $T_2$ time is always pursued. To date, more than 30 s $T_2$ has been achieved for optically trapped single strontium atoms \cite{kaufman_30s_clock,atomcomputing_nuclear_qubit}. The coherence time is obtained by either using optical tweezers with a ``magic wavelength'' \cite{kaufman_30s_clock}, where the differential light shift (DLS) between the two clock states can be cancelled, or nuclear spins \cite{atomcomputing_nuclear_qubit}. Thus, coherence is intrinsically immune to fluctuations in the trapping light intensity.

However, for widely used alkali metal atoms, such as rubidium (Rb) atoms and cesium (Cs) atoms, the $T_2$ time of the qubit encoded in the microwave clock states of optically trapped single atoms is much shorter due to the lack of a ``magic wavelength''. The DLS between two clock states is susceptible to fluctuations in the trapping light intensity \cite{Kuhr2005}. In recent years, many efforts have been made to improve the $T_2$ time of optically trapped alkali metal atoms by finding other ``magic conditions'' \cite{Li2019, Yang2016, Carr2016, Guo2020, Kazakov2015, Sarkany2014, Kim2013, Derevianko2010, Chicireanu2011, Derevianko2011, Jai2007}, where the first-order dependence of the DLS on fluctuations in light intensity and/or magnetic fields could be suppressed. By applying such ``magic conditions'', the $T_2$ time can be improved from tens of milliseconds to the second level. 

Another commonly adopted method for obtaining a long coherence time is to decrease the temperature of the trapped atom. In a full quantum picture where the vibrational quantum states (phonon states) are taken into account \cite{PFYang_2022}, a trapped single atom with a higher temperature has a wider population distribution on the phonon states. Due to the difference in the DLS associated with the different phonon states, the Ramsey signal is the sum of the interfering signals between the two atomic internal electronic states associated with different phonon states, and a fast decay of the amplitude is observed, which is usually referred to as inhomogeneous decoherence. The inhomogeneous decoherence can be recovered by using a spin-echo technique \cite{PFYang_2022, Kuhr2005}. Thus, the usually assumed residual decoherence factors \cite{Kuhr2005}, such as fluctuations in the trap depth, trap position, magnetic field, etc., will no longer depend on the atomic temperature. Moreover, without using ``magic conditions'', a $T_2$ time of 12.6 s for single cesium atoms in blue-detuned traps was recently obtained \cite{Weiss_SGdetection} by adopting Carr-Purcell-Meiboom-Gill (CPMG) pulse sequences with the atom cooled to its ground phonon state. However, the decoherence mechanism associated with the atomic temperature has not been revealed.  

In this article, we propose a mechanism of decoherence associated with the atomic temperature. This decoherence is caused by the stochastic jump of the atomic phonon state due to trapping noise.
We therefore name this decoherence mechanism phonon-jumping-induced decoherence (PJID). 
To understand the PJID between the two internal electronic states $|a\rangle$ and $|b\rangle$, we have to take the external vibrational quantum states (phonon states) of the atom in optical traps into account of the full quantum wavefunction. 
The motion of the trapped atom in trap potentials caused by the light shifts of electronic states $|a\rangle$ and $|b\rangle$ is described by three-dimensional (3D) quantum harmonic oscillators.
The motional evolution and the phonon jumping caused by the noise of the trapping field in the two trap potentials are independent because there is no interaction between the two internal states when the atom is freely evolving. 
Stochastic phonon jumping of the atom will induce decoherence between the two internal states $|a\rangle$ and $|b\rangle$.
The PJID mechanism results in an exponential decay of coherence, which differs from the Gaussian decay of decoherence caused by the variance of the DLS. These two mechanisms occur simultaneously in the decoherence process. 
By examining the decoherence process of a Cs atom in a red-detuned optical trap with different intensity noise levels, we introduce PJID and show that experimental data is consistent with our model for PJID.
After fully understanding the decoherence mechanisms, we adopt a blue-detuned BBT and cool the atom in its 3D motional ground states to suppress both the PJID and variance of the DLS.
A coherent time $T_2$ of approximately 20 seconds for a qubit encoded in the clock states of a single trapped Cs atom is finally obtained. To the best of our knowledge, this is the longest coherence time for a qubit encoded in an optically trapped single alkali metal atom, and it can be improved further by improving the phase noise of the driving microwave, the pointing noise of the trap, etc.

\section{The phonon-jumping-induced decoherence}

\begin{figure}%[htbp]
\centering
\includegraphics[width=\columnwidth]{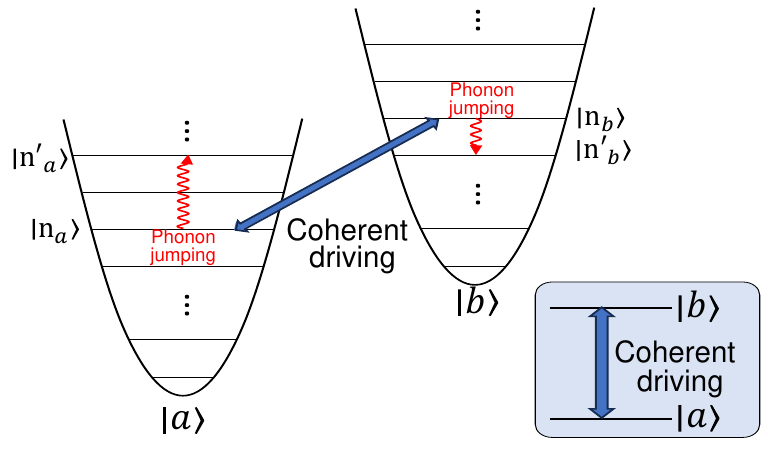}
\caption{\label{fig1}
Principle of PJID. The atom oscillates in the two optical traps produced by light shifts of the electronic states $|a\rangle$ and $|b\rangle$. The motional evolutions and the phonon jumpings caused by the noises of trapping field in the two trap potentials are independent because there is no interaction between the two states when the atom is freely evolving. The phonon jumps induced by the noise of the trap light then destroy the coherence between the two electronic states. The coherence is usually characterized by the interference of the two states with the aid of coherent driving between the two electronic states with the same phonon number.
} 
\end{figure}

As shown in Fig. \ref{fig1}, the atom vibrational states (phonon states) are denoted by $|n_q\rangle$, where $n_q$ is the phonon number (PN) along the vibrational axis $q$ ($q=x$, $y$, or $z$). 
The phonon states obey the orthogonal relation $\langle n_q | n'_q \rangle = \delta_{n_q,n'_q}$. In a rotating frame associated with the atom frequency between the qubit states $|a\rangle$ and $|b\rangle$, the time-dependent full wavefunctions of the atomic qubit can be expressed as $|\psi_{a}(t)\rangle=|a \rangle \otimes \prod^{\otimes}_q |n_{q,{a}}\rangle$ and $|\psi_{b}(t)\rangle=\exp{[-i(\Delta^\text{DLS} t+\phi)]} |b\rangle \otimes \prod^{\otimes}_q |n_{q,{b}}\rangle $. Here, 
$\Delta^\text{DLS} =-\eta \frac{U_0}{\hbar}+ \frac{\eta}{2} \sum_q (n_q+\frac{1}{2}) \omega_q$
is the DLS between states $|a\rangle \otimes \prod^{\otimes}_q |n_{q,{a}}\rangle$ and $|b\rangle \otimes \prod^{\otimes}_q |n_{q,{b}}\rangle$ \cite{PFYang_2022}, and $\phi$ is the additional phase. $\eta=\left| \frac{\omega_\text{hfs}}{\Delta} \right|$ is the ratio between the hyperfine splitting and the frequency detuning of the trap light. $U_0$ is the potential at the trap center, and $\omega_q$ is the oscillation frequency. The coherence between the two states is
\begin{align}
\text{C}(t)=\int \text{d}\Delta^\text{DLS} \int \text{d} \phi f(\Delta^\text{DLS}) \varphi(\phi) \text{Tr}(|\psi_{a}(t)\rangle \langle \psi_{b}(t)|), \label{decoh}
\end{align}
where $f(\Delta^\text{DLS})$ and $\varphi(\phi)$ are the probability distributions of $\Delta^\text{DLS}$ and $\phi$, respectively. The trace is made over both the electronic and phonon state spaces. Then, it is rewritten as
\begin{equation}
\begin{split}
&\text{C}(t)=\\
&\int \text{d}\Delta^\text{DLS} \int \text{d} \phi f(\Delta^\text{DLS}) \varphi(\phi) \textbf{e}^{[-i(\Delta^\text{DLS} t+\phi)]} \prod_q \delta_{n_{q,a},n_{q,b}} . \label{coh}
\end{split}
\end{equation}

We first discuss the DLS-dependent part of Eq. (\ref{coh}), which can be separated as 
\begin{equation}
\text{C}_1 (t)=\int \text{d}\Delta^\text{DLS} f(\Delta^\text{DLS}) \textbf{e}^{-i \Delta^\text{DLS} t} . \label{coh2}
\end{equation}

For a given spatial structure of the optical trap, the potential $U_0$ and trap frequency $\omega_q$ are determined by the total trap power $P$. The probability distribution of $P$ usually follows a Gaussian function $\exp{[-(P-P_0)^2/(2\sigma_P^2)]}$ with $P_0$ and $\sigma_P$ as the mean and root mean square (rms) values of the total trap power, respectively. The probability distribution of $\Delta^\text{DLS}$ also follows a Gaussian function with $f(\Delta^\text{DLS}) \propto \exp{[-(\Delta^\text{DLS}-\Delta^\text{DLS}_0)^2/(2\sigma_\text{DLS}^2)]}$, where $\Delta^\text{DLS}_0$ is the mean value of DLS. The variation in the DLS ($\sigma_\text{DLS}$) depends on $\sigma_P$ by
\begin{equation} \label{DLS-var}
\sigma_\text{DLS} =-\eta \frac{U_0}{\hbar} \frac{\sigma_P}{P_0} + \frac{\eta}{4} \sum_q (n_q+\frac{1}{2}) \omega_q  \frac{\sigma_P}{P_0}.
\end{equation}

By setting $\text{C}_1(0)=1$, the decay of the coherence given in Eq. (\ref{coh2}) will finally take a Gaussian form
\begin{equation} \label{coh3}
 \text{C}_1(t)=\textbf{e}^{-\sigma^2_\text{DLS} t^2 /2}. 
\end{equation}

Next, we will discuss the rest of Eq. (\ref{coh}), which is connected to the stochastic phonon jumps induced by the noise of the trap light. As a consequence, the jump of either the PN or the phase would cause decoherence. 
We assume that the atom is initially prepared in a superimposed state of $|a\rangle \otimes \prod^{\otimes}_q |n_{q,{a}}\rangle$ and $|b\rangle \otimes \prod^{\otimes}_q |n_{q,{b}}\rangle$ with $n_{q,{a}}=n_{q,{b}}$ by a coherent driving field. Assuming that the PN associated with one electronic state alternates due to the noise at time $t$, according to Eq. (\ref{coh}), the coherence immediately collapses because $\delta_{n'_{q,a},n'_{q,b}}=0$ due to $n'_{q,{a}} \neq n'_{q,{b}}$. Here, $n'_{q,{a}}$ and $n'_{q,{b}}$ are the PNs at time $t$. Even if the PNs are alternated simultaneously to the same number ($n'_{q,{a}} = n'_{q,{b}}$) at time $t$, the coherence also disappears due to the stochasticity of the noise-induced phonon jumping process. In this case, the phase $\phi$ in Eq. (\ref{coh}) is evenly distributed in the range [0, 2$\pi$) with $\varphi(\phi)=1/2\pi$. Therefore, $\int^{2\pi}_{0} \text{d} \phi \varphi(\phi)=0$ and the coherence is $\text{C}=0$.

The process of decoherence is determined by the jumping rate of the PN. If we define the jumping rate from PN $n_q$ along axis $q$ as $R_{q}$, the probability of the atom being in state $|n_q\rangle$ is $p_{n_q}$. Then, $p_{n_q}$ obeys the rate equation $\dot{p}_{n_q}=-R_{q} p_{n_q}$. 
Hence, the coherence takes the form 
\begin{equation}  \label{coh4}
 \text{C}_2(t)=\textbf{e}^{-(R_x +R_y + R_z) t }.
\end{equation}
In an optical trap, phonon jumping is induced by the intensity noise and the beam-pointing noise associated with the trap light \cite{Savard1997,Gehm1998}. The intensity (beam-pointing) noise will cause the PN to jump by two (one). The overall phonon jumping rate (PJR) from state $|n_q\rangle$ is the sum of all the jumping rates given in \cite{Savard1997,Gehm1998}, and the result is
\begin{equation}
R_{q}=\frac{\pi \omega_q^2}{8} S_k (2 \omega_q) ((n_q+1)^2-n_q) + \frac{\pi}{2 \hbar} M \omega_q^3 S_q(\omega_q) (2n_q+1), \label{rate3}
\end{equation}
where $M$ is the mass of the trapped atom and $\omega_q$ is the trap frequency along the $q$ axis. $S_k(\omega)$ and $S_q(\omega)$ are the one-sided power spectra of the fractional fluctuations in the spring constant and coordinate $q$.  

Finally, we obtain the decay of the overall coherence 
\begin{equation}  \label{coh5}
 \text{C}(t)=\text{C}_1(t) \text{C}_2(t)=\textbf{e}^{-\sigma^2_\text{DLS} t^2 /2- R t},
\end{equation} where $R=R_x +R_y + R_z$ is the overall phonon jumping rate (PJR). We see that the coherence actually shows a Gaussian and exponential combined decay. It should be noted that the above analysis are based on a atom with specific phonon number $n_q$. For a thermal atom the overall variation in the DLS and PJR should be obtained by $ \sigma^2_\text{DLS, all}=\sum_{n_q} P({n_q}) \sigma^2_\text{DLS}$ and $R_\text{all}=\sum_{n_q} P({n_q}) R$ with $P({n_q}) = \langle n_q\rangle^{n_q}/ (\langle n_q\rangle+1)^{n_q+1}$ the probability of atom on phonon state $|n_q\rangle$ and $\langle n_q\rangle$ the mean phonon number. 

% \section{Experimental validation of PJID}
\begin{figure}%[htbp]
\centering
\includegraphics[width=\columnwidth]{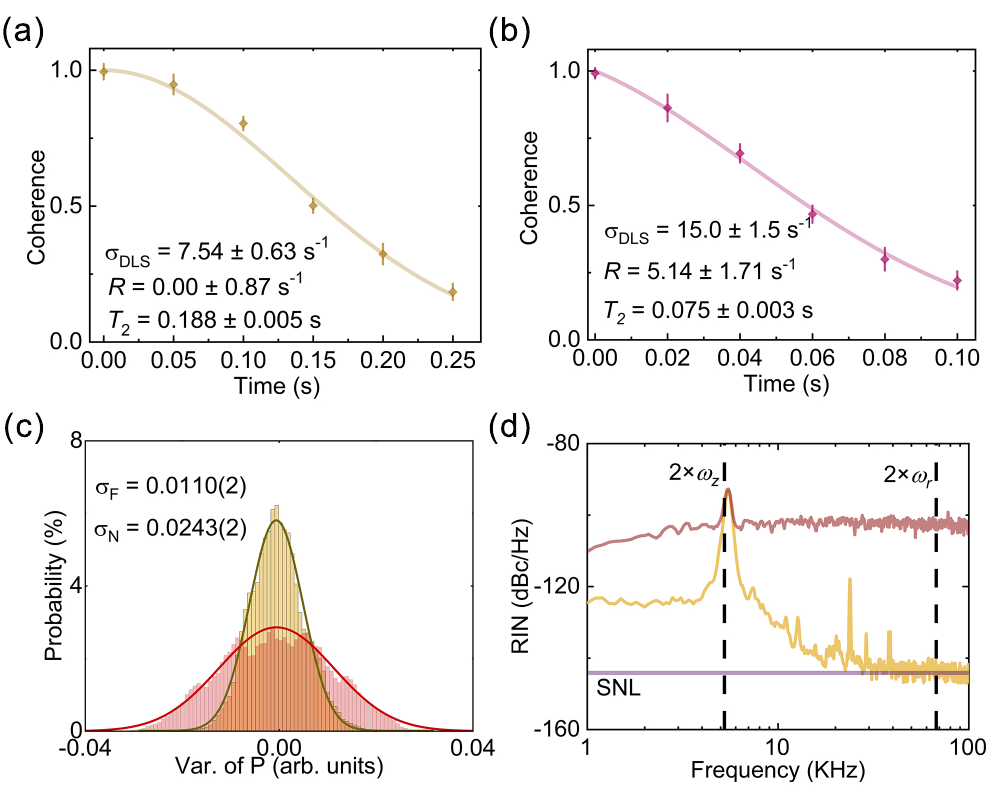}
\caption{\label{fig2}
Coherence decay of a single Cs atom trapped in a 1052-nm ODT. (a) and (b) are the coherence decays at the conditions that the laser is free running (low intensity noise) and 40-dB intensity noise is added. (c) displays the distributions of the sampled light power, which is normalized to the mean value, for the two conditions. $\sigma_F$ and $\sigma_N$ are the normalized power variances for the free-running laser and laser with 40-dB noise added, respectively. (d) displays the noise spectra, in which the dashed lines marked as $2\times \omega_z$ and $2\times \omega_r$ are the frequencies where the parametric process occurs. SNL: shot-noise limit.
} 
\end{figure}

In the experiment of optically trapped single atoms, a trapping laser with low intensity noise and pointing noise is usually adopted to get longer atom storage time and coherence time. In such situation, the PJID is much smaller than the DLS variance. To experimentally check the effect of PJID, we deliberately amplify the PJID by adding a 40-dB intensity noise on the trap light. We measured two coherence decays in a red-detuned ODT with a normal intensity noise and a deliberately enlarged intensity noise. It should be noted that the 40-dB noise will not affect the state lifetime $T_1$, but the atom lifetime is reduced to approximate 1 s compare to the 30 s lifetime without the noise. The ODT is formed by strongly focusing a 1052-nm laser beam to a size of 1.65 $\mu$m and loading single Cs atoms from a magneto-optical trap (MOT). The coherence between the clock states ($|6S_{1/2} F=3, m_F=0 \rangle$ and $|6S_{1/2} F=4, m_F=0 \rangle$) was measured by a standard spin-echo interferometer \cite{Hahn1950,Andersen_Echo} with a 9.2-GHz microwave driving field. Figure \ref{fig2}(a) and (b) shows the coherence data for different time delays under the condition that the trap laser is free running and 40-dB intensity noise is added in a frequency range that covers the trap frequencies. In Fig. \ref{fig2}(a), the coherence decays more like a Gaussian function because of the low PJR. The data fitting by Eq. (\ref{coh5}) gives a DLS variation $\sigma_\text{DLS}= 7.54 \pm 0.63$ s$^{-1}$ and a PJR $R= 0.00 \pm 0.87$ s$^{-1}$. The corresponding $T_2=188 \pm 5 $ ms, which is defined by $1/e$ of the coherence. However, the coherence data in Fig. \ref{fig2}(b), where 40-dB intensity noise is added to the trap light, apparently deviate from the Gaussian function, and the $1/e$ coherence time is $T_2=75 \pm 3$ ms. In this situation, Eq. (\ref{coh5}) provides good data fitting, and the fitted DLS variation and PJR are $\sigma_\text{DLS}= 15.0 \pm 1.5$ s$^{-1}$ and $R= 5.14 \pm 1.71$ s$^{-1}$, respectively.
Compared to the condition in which the laser is free running, the DLS variation is increased by a factor of two, which is in agreement with the increase ($\approx 2.2$) in the variance of the trap light intensity [Fig. \ref{fig2}(c)]. The PJR is increased by 5.14 s$^{-1}$, which comes from the parametric process-induced phonon jumping [the first term in Eq. (\ref{rate3})] because the second term remains the same in the two situations. By using the measured intensity noise [Fig. \ref{fig2}(d)], the increase in the PJR can be estimated as 6.0 s$^{-1}$ \cite{SM}, which agrees well with the number obtained by the data fitting. Therefore, the existence of PJID can be confirmed.

\section{20-s coherence time of a qubit encoded in an optically trapped single Cs atom}
A long coherence time can be obtained by suppressing both the DLS variance and PJR. The DLS variance can be greatly suppressed by adopting red-detuned optical traps with ``magic conditions'' \cite{Li2019}. However, because the atom is confined in the region of intensity maxima, the decoherence induced by photon scattering is also maximized \cite{SM}. In a well-aligned blue-detuned trap, the atom is trapped in the region with zero light intensity in principle. Thus, the decoherence induced by photon scattering is completely canceled. The main term  of DLS [the first term on the right-hand side of Eq. (\ref{DLS-var})] also disappears. The rest of the DLS due to the phonon energy can be suppressed by preparing the atom in its three-dimensional (3D) motional ground state (zero phonon state, ZPS), i.e., $n_q=0$.

As given by Eq. (\ref{rate3}), the PJR is determined by the trap frequency $\omega_q$, intensity noise $S_k$, pointing noise $S_q$, and PN $n_q$. 
Therefore, it can be suppressed by reducing these parameters. 
The trap frequency $\omega_q$ is determined by the depth and size of the optical trap and thus can be reduced by using a trap with shallow depth and large size.
The intensity noise can be suppressed by applying a noise eater or adopting a low-noise laser.
The pointing noise can be minimized by improving the mechanical stability of the trap optics.
The most efficient way to suppress the PJR is to decrease the PN. PJID can be minimized by preparing the atom in its 3D ZPS.

\begin{figure}%[htbp]
\centering
\includegraphics[width=\columnwidth]{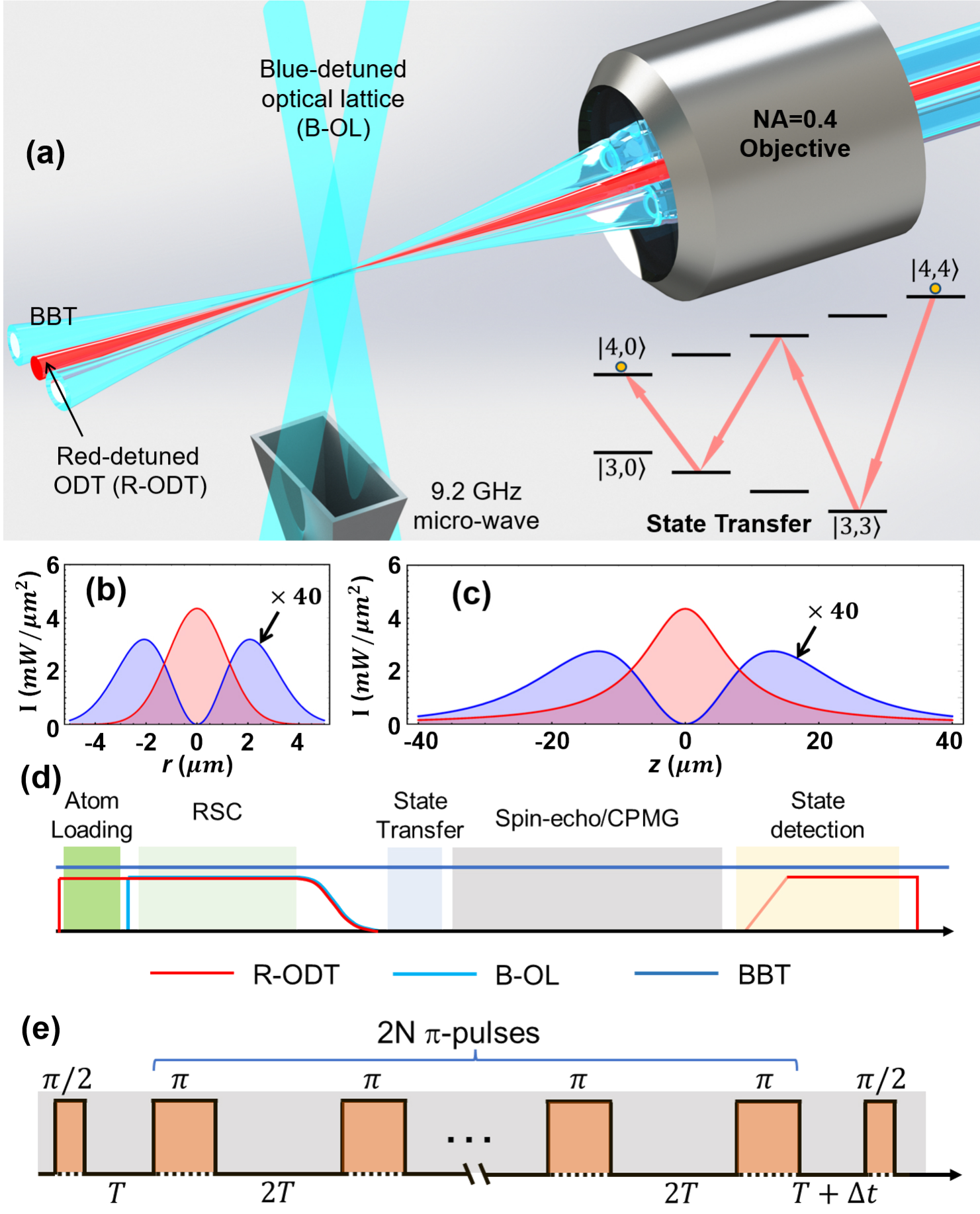}
\caption{\label{fig3}
(a) Schematic experimental setup for the long coherence time of a single Cs atom in a BBT. The R-ODT trap is used to load a single atom. The R-ODT and B-OL combined trap is used for the motional ground state cooling of the atom by Raman sideband cooling. The state is transferred from $|6S_{1/2} F=4, m_F=4 \rangle$ to $|6S_{1/2} F=4, m_F=0 \rangle$ after the RSC by four $\pi$-pulses. (b) and (c) display the intensity profiles of the R-ODT and BBT along the radial and the axial directions, respectively. The BBT laser intensity is plotted with 40 times enlargement. (d) The time sequence for a single iteration of the experiment. (e) The Carr-Purcell-Meiboom-Gill pulse sequence used for suppressing the residual DLS variance.}
\end{figure}

\begin{figure}%[htbp]
\centering
\includegraphics[width=\columnwidth]{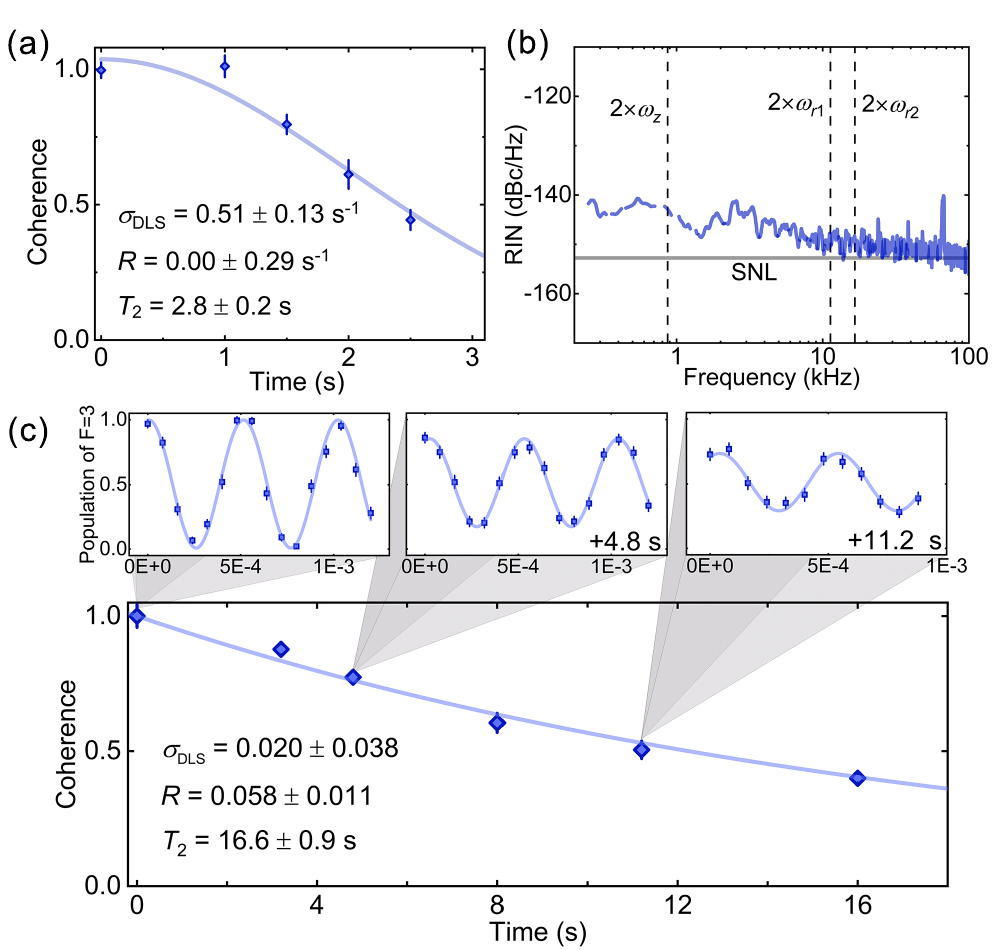}
\caption{\label{fig4}
Coherence decay of a single Cs atom in the 780-nm BBT. (a) is the coherence decays obtained by spin-echo spectroscopy. (b) is the relative intensity noise of the BBT trap light, and the dashed lines are the frequencies at which the parametric process occurs. SNL: shot-noise limit. (c) is the coherence decay by CPMG sequence. Three interference fringes at time delays of 0, 4.8, and 11.2 s obtained by the CPMG sequence are also shown as insets. }
\end{figure}

Here, we adopt a blue-detuned optical bottle beam trap (BBT) to demonstrate the long coherence of a qubit encoded in a single Cs atom by suppressing both the DLS variance and the PJR. The trap is formed by focusing two parallelly propagating 780-nm vortex laser beams with orthogonal polarization by a single objective with a numerical aperture $NA=0.4$ [Fig. \ref{fig3}(a)]. 
Therefore, the two beams cross each other around the focus of the objective, and a microsized BBT is formed \cite{Li2012}. 
The radius and length of the BBT are $r=2.9$ $\mu$m and $l=14.3$ $\mu$m, respectively.
The details of the trap construction can be found in the Supplementary Materials \cite{SM}.
The intensity ratio between the trap center and the trap barrier is 1.5\%. 
Thus, the DLS variance can be greatly reduced by using a shallow trap depth and preparing the atom in its motional 3D ground states. The PJR can also be suppressed by adopting a low intensity noise laser and improving the mechanical stability of the trap. 
We use a 780-nm external-cavity diode laser to build the trap. The output has very low intensity noise [Fig. \ref{fig4}(b)], and the relative intensity variance is smaller than 0.015\%.
By using a trap power of 9 mW, we can build a trap with a minimum barrier of $k_B \times 50 $ $\mu$K ($k_B$ is the Boltzmann constant). The trap frequencies are $(\omega_x,\omega_y,\omega_z)=2 \pi \times (5.65,8.3,0.435)$ kHz.
For a Cs atom in its 3D ZPS, the estimated DLS variance and the PJR due to the intensity noise are $\sigma_\text{DLS}<3.0\times 10^{-3}$  s$^{-1}$ and $R_k<2.5\times 10^{-6}$ s$^{-1}$, respectively. The pointing noise of the BBT is difficult to measure, and it is improved by mounting all the optics with the cage system from Thorlabs. With the atom in its 3D ZPS, the PJR due to pointing noise can also be minimized.
Therefore, a long $T_2$ time is expected.

The single atom in its 3D ZPS is prepared with the aid of a combined microsized ODT. Due to the loose confinement of the atom in the BBT, it is impossible to directly cool the atom to its 3D ZPS. Therefore, an extra combined microsized ODT with strong confinement is adopted. The experimental sketch is displayed in Fig. \ref{fig3}(a), and the BBT and ODT are arranged to be overlapped in both the radial and axial directions [Fig.\ref{fig3}(b) and (c)]. The time sequence is shown in Fig.\ref{fig3}(d) and the experimental details can be found in SM \cite{SM}.  Resolved Raman sideband cooling (RSC) is used to prepare the atoms in 3D ZPS, and the residual phonon numbers are $(\bar{n}_x,\bar{n}_{y},\bar{n}_z) \simeq (0.07\pm0.07,0.04\pm0.04,0.08\pm0.06)$ \cite{Tian2023_RSC}. The atom populates on state $|6S_{1/2} F=4, m_F=4 \rangle$ after the RSC process. The combined ODT is then adiabatically turned off, and the atom is resettled in the BBT without changing its phonon state and electronic state. The atom is transferred to $|6S_{1/2} F=4, m_F=0 \rangle$ by four microwave $\pi$-pulses via the intermediate states $|6S_{1/2} F=3, m_F=3 \rangle$, $|6S_{1/2} F=4, m_F=2 \rangle$, and $|6S_{1/2} F=3, m_F=1 \rangle$. A spin-echo between the Cs clock states ($|6S_{1/2} F=3, m_F=0 \rangle$ and $|6S_{1/2} F=4, m_F=0 \rangle$) is performed to evaluate the decay of the coherence.

The coherence decay measured by a spin-echo interferometer is displayed in Fig. \ref{fig4}(a). The data show a Gaussian function decay, and the fitting by Eq. (\ref{coh5}) gives the DLS variance $\sigma_\text{DLS}=0.51 \pm 0.13$ s$^{-1}$ and PJR $R=0.00 \pm 0.29$ s$^{-1}$. The variation of DLS dominates the decay, thus the fitted $R$ has a very large error and is inaccurate. The corresponding coherence time is $T_2=2.8 \pm 0.2$ s. The DLS variance is much larger than we estimated. This is probably due to the slow disturbance of the energy levels caused by the variations in the magnetic field and the BBT light field which are not taken into account in the discussions above. A CPMG pulse sequence is then applied to decouple the spin dynamics from these slow disturbances \cite{Dynamical_decoupling,PhysRevLett.106.240501}. The pulse sequence is shown in Fig. \ref{fig3}(e), where a series of $\pi$-pulses are inserted between the two $\pi/2$-pulses. The time interval $T=0.8$ s is chosen for its best performance. The CPMG pulse sequence works as a filter that can filter out slow DLS variations with frequencies $n\times 0.0625$ Hz ($n=1,2,3,\cdots$). The obtained decay of coherence turns out to be much slower. The data are shown in Fig. \ref{fig4}(c). The residual DLS variance and PJR are fitted by Eq. (\ref{coh5}) with $\sigma_\text{DLS}=0.020 \pm 0.038$ s$^{-1}$ and $R=0.058 \pm 0.011$ s$^{-1}$. The variation of DLS $\sigma_\text{DLS}$ is dramatically suppressed, and the PJID then dominates the coherence decay. The small DLS variance might come from residual variations in the magnetic field and BBT light. The stability and phase noise of the microwave source may also contribute to the DLS variance. The main source of the residual PJR should be the pointing noise of the BBT. All of these factors could be suppressed further by using proper methods.
In the current case, the coherence time is $T_2=16.6\pm 0.9$ s. Accounting for the single atom lifetime of $105.5 \pm 13.1$ s \cite{SM}, the actual coherence time should be $T_2=19.7$ s.

\section{Conclusion}
We discovered and experimentally verified a new decoherence mechanism, i.e., phonon-jumping-induced decoherence, for a single-alkali-atom qubit trapped in an optical dipole trap. Then, a coherence time of approximately 20 seconds is obtained for a qubit encoded in a single Cs atom by suppressing both the DLS variance and the PJID by adopting a blue-detuned BBT and preparing the atom into its 3D ZPS. This is the longest coherence time reported to date for a qubit encoded in an optically trapped alkali metal atom. Our analysis and the results and corresponding methods for improving coherence are universal and can be applied to other atoms and molecules, opening up new prospects for expanding the coherent manipulation of optically trapped atoms or molecules.

\begin{acknowledgements}
This work was supported by the National Key Research and Development Program of China (Grant No. 2021YFA1402002), Innovation Program for Quantum Science and Technology (Grant No. 2023ZD0300400), the National Natural Science Foundation of China (Grant Nos. U21A6006, U21A20433, 11974223, 11974225, 12104277, and 12104278), the Fund for Shanxi 1331 Project Key Subjects Construction, and Fundamental Research Program of Shanxi Province (202203021223003).
\end{acknowledgements}

\bibliographystyle{Zou}
\bibliography{decoherence-BlueTrap}

\pagebreak

\clearpage

\widetext

\setcounter{page}{1}

\renewcommand{\thefigure}{S\arabic{figure}}
\setcounter{figure}{0}
\renewcommand{\thetable}{S\arabic{table}}
\setcounter{table}{0}
\renewcommand{\theequation}{S\arabic{equation}}
\setcounter{equation}{0}

\section*{Supplementary Materials for \\
``Extending the coherence time limit of a single-alkali-atom qubit by suppressing phonon-jumping-induced decoherence''}

\subsection*{Zhuangzhuang Tian, Haobo Chang, Xin Lv, Mengna Yang, Zhihui Wang, Pengfei Yang, Pengfei Zhang, Gang Li, and Tiancai Zhang}

\subsection*{State Key Laboratory of Quantum Optics and Quantum Optics Devices,
and Institute of Opto-Electronics, Shanxi University, Taiyuan 030006, China\\
Collaborative Innovation Center of Extreme Optics, Shanxi University, Taiyuan 030006, China
}

The supplementary materials include the calculation of phonon jumping rate, experiment details of the single Cs atom in the 1052-nm ODT and the 780-nm BBT, and analysis of photon-scattering induced decoherence.

\section{The total phonon jumping rate}

In the conventional adopted optical trap, the motion of an atom can be treated as a three-dimensional (3D) quantum harmonic oscillator (HO) when the temperature of the atom is much lower than the trap depth. The jump of the trapped particle between different phonon states is caused by fluctuations in the trap potential, where two mechanisms are included: fluctuations in the spring constant and fluctuations in the trap position. The induced jumping rates between different phonon states with a one-dimensional (1D) HO are given in Refs. \cite{Savard1997,Gehm1998}
\begin{equation}
R_{n \pm 2 \leftarrow n}= \frac{\pi \omega^2}{16} S_k (2 \omega)  (n+1 \pm 1)(n\pm1) \label{rate1}
\tag{S1}
\end{equation}
and 
\begin{equation}
R_{n \pm 1 \leftarrow n}= \frac{\pi}{2 \hbar} M \omega^3 S_q(\omega) (n+1/2 \pm 1/2), \label{rate2}
\tag{S2}
\end{equation}
where $M$ is the mass of the trapped particle and $\omega$ is the trap frequency. $S_k(\omega)$ and $S_q(\omega)$ are the one-sided power spectra of the fractional fluctuations in the spring constant and coordinate $q$, respectively, which are determined by the intensity and pointing noise of the optical trap beam. The overall jumping rate of the phonon number (PN) from $n$ is the sum of all the jumping rates given by Eqs. (\ref{rate1}) and (\ref{rate2}).
\begin{equation}
R_n = \frac{\pi \omega^2}{8} S_k (2 \omega) ((n+1)^2-n) + \frac{\pi}{2 \hbar} M \omega^3 S_q(\omega) (2n+1). \label{HITRate}
\tag{S3}
\end{equation}

For a thermal atom trapped in an optical dipole trap as we used to check the existence of the PJID in Section 2 of the main text, the PJR can be estimated by the mean energy of the atom. According to the equipartition theorem, for a thermal atom with temperature is $T$, the mean kinetic energy of the atom on every axis is $(\langle n_q \rangle+\frac{1}{2}) \hbar \omega_q =k_B T/2$, where $\langle n_q \rangle$ is the average photon number. The phonon jumping rate in Eq. (\ref{HITRate}) can be approximated by
\begin{equation}
R_{\langle n_q \rangle}\approx\frac{\pi}{8 \hbar^2} S_k (2 \omega_q) (k_B T / 2)^2 + \frac{\pi}{2\hbar^2} M \omega_q^2 S_x(\omega_q) k_B T. \label{HITRate-T}
\tag{S4}
\end{equation}
The overall phonon jumping rate can be calculated by summing the phonon jumping rates on the three axes.
\begin{equation}
    \begin{split}
            R_\text{PJR}=&\frac{\pi}{8 \hbar^2} (k_B T / 2)^2 \sum_{q=x,y,z} S_k (2 \omega_q)  \\
                 & + \frac{\pi}{2 \hbar^2} M k_B T \sum_{q=x,y,z} \omega^2_q S_q(\omega_q) . \label{HITRate-TA}
    \end{split}
    \tag{S5}
\end{equation}
The phonon jumping rates of single Cs atoms due to the parametric process in the red-detuned ODT or BBT are also estimated by the first term on the right hand side of this equation. The relation $S_k (\omega)=S_k (f)/(2\pi)$ is also adopted in the estimation.

\section{Experimental details of the single Cs atom in the 1052-nm ODT}

\begin{figure}[t]
\centering
\includegraphics[width=0.5\columnwidth]{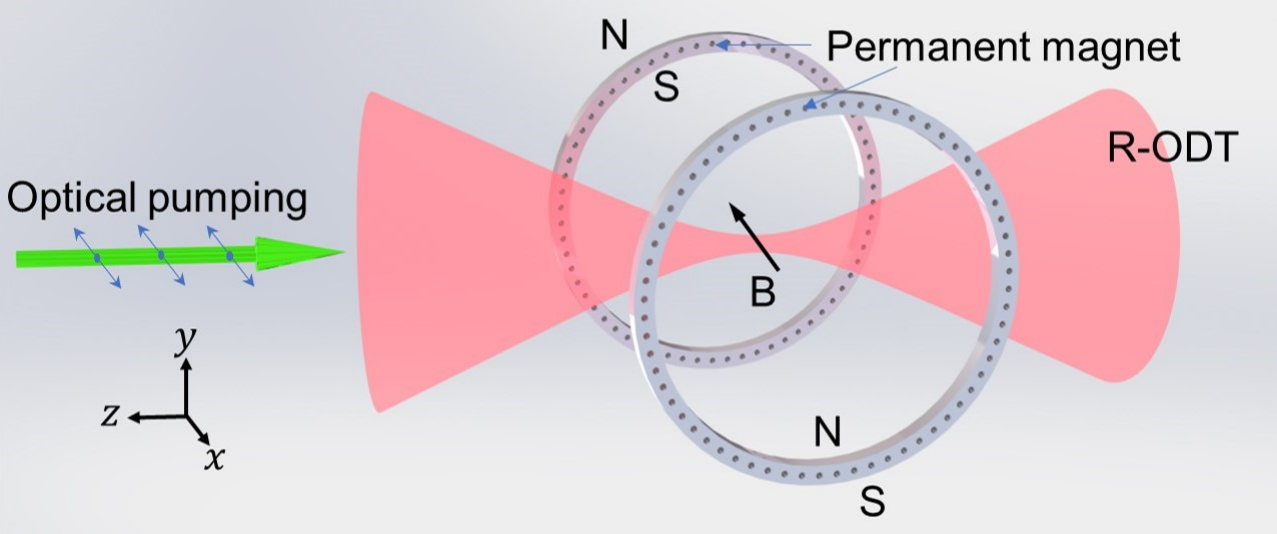}
\caption{The experimental sketch for single atom manipulation in the red-detuned optical dipole trap (R-ODT). A magnetic field with B = 1.7 Gauss generated by permanent magnets is used to serve as a quantization axis. A $\pi$-polarized optical pumping beam (green arrow), which is resonant with $|6S_{1/2}, F=4\rangle \leftrightarrow |6P_{1/2}, F'=4\rangle$, and the MOT repumping light (resonant with $|6S_{1/2}, F=3\rangle \leftrightarrow |6P_{3/2}, F'=4\rangle$, not shown in the figure) are used to initialize the atom to state $|6S_{1/2}, F=4, m_F=0 \rangle$.}
\label{RODT}
\end{figure}

An experimental sketch of single atom manipulation in a red-detuned optical dipole trap (R-ODT) is shown in Fig. \ref{RODT}. The R-ODT is obtained by focusing a 1052-nm laser beam with an NA = 0.4 objective. The $1/e^2$ beam radius is 1.65 $\mu$m, which is inferred from the trap frequency. We first load a single atom from cold atomic ensembles obtained by a magneto-optical trap (MOT). The trapped single atom is then cooled by polarization gradient cooling (PGC) to a temperature of approximately 10$\mu$K. A magnetic field with B = 1.7 Gauss generated by permanent magnets is used to serve as a quantization axis. Next, the atom is initialized to state $|F=4, m_F=0\rangle$ by a combination of a $\pi$-polarized optical pumping beam, which is resonant with $|6S_{1/2}, F=4\rangle \leftrightarrow |6P_{1/2}, F'=4\rangle$, and the MOT repumping light, which is resonant with $|6S_{1/2}, F=3\rangle \leftrightarrow |6P_{3/2}, F'=4\rangle$. A Ramsey or spin-echo interferometer is then applied by a 9.2-GHz microwave which drives the clock transition ($|6S_{1/2}, F=3, m_F=0 \rangle \leftrightarrow |6S_{1/2}, F=4, m_F=0 \rangle$). The Rabi frequency of clock transition is 33.7 kHz, so that a $\pi/2$-pulse length is 7.4 $\mu$s and a $\pi$-pulse length is 14.8 $\mu$s. The atom state is ultimately detected by counting the atom events after blowing away the atom in state $|6S_{1/2}, F=4\rangle$ with a light resonance with $|6S_{1/2}, F=4\rangle \leftrightarrow |6P_{3/2}, F'=5\rangle$. 

\begin{figure}[t]
\centering
\includegraphics[width=0.5\columnwidth]{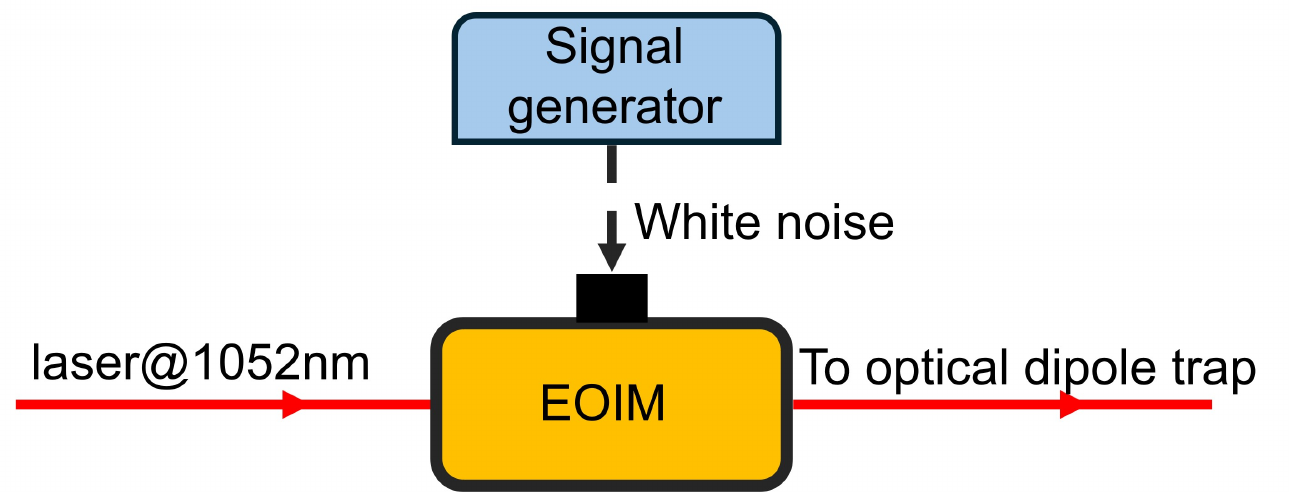}
\caption{Schematic of applying the additional intensity noise to the trap laser. The electronic white noise generated by the signal generator modulates the intensity of the trap laser via the electro-optic intensity modulator (EOIM).}
\label{noise}
\end{figure}

The schematic for the laser intensity noise control is shown in Fig. \ref{noise}. A signal generator (MODEL DS345, Stanford Research Systems) is used to generate white noise with a bandwidth of 10 MHz and an amplitude of 9 Vpp. The white noise is fed into an electro-optic intensity modulator (EOIM, EO-AM-NR-C2, Thorlabs) to applied the noise on the laser light. By changing the parameters of the signal generator, the intensity noise of dipole trap laser can be controlled.
The intensity noise of the laser is measured by a photodetector (PDA05CF2, Thorlabs) with a bandwidth of 150 MHz, and the noise spectrum is analyzed by a spectrum analyzer ($RBW= 300$ Hz). The temporal variance is measured by an oscilloscope with sampling rate of 2.5 MHz and the sampling time of 4 seconds. The results are shown as Fig. 2 (c) and (d) in main text.

\begin{figure}[t]
\centering
\includegraphics[width=0.5\columnwidth]{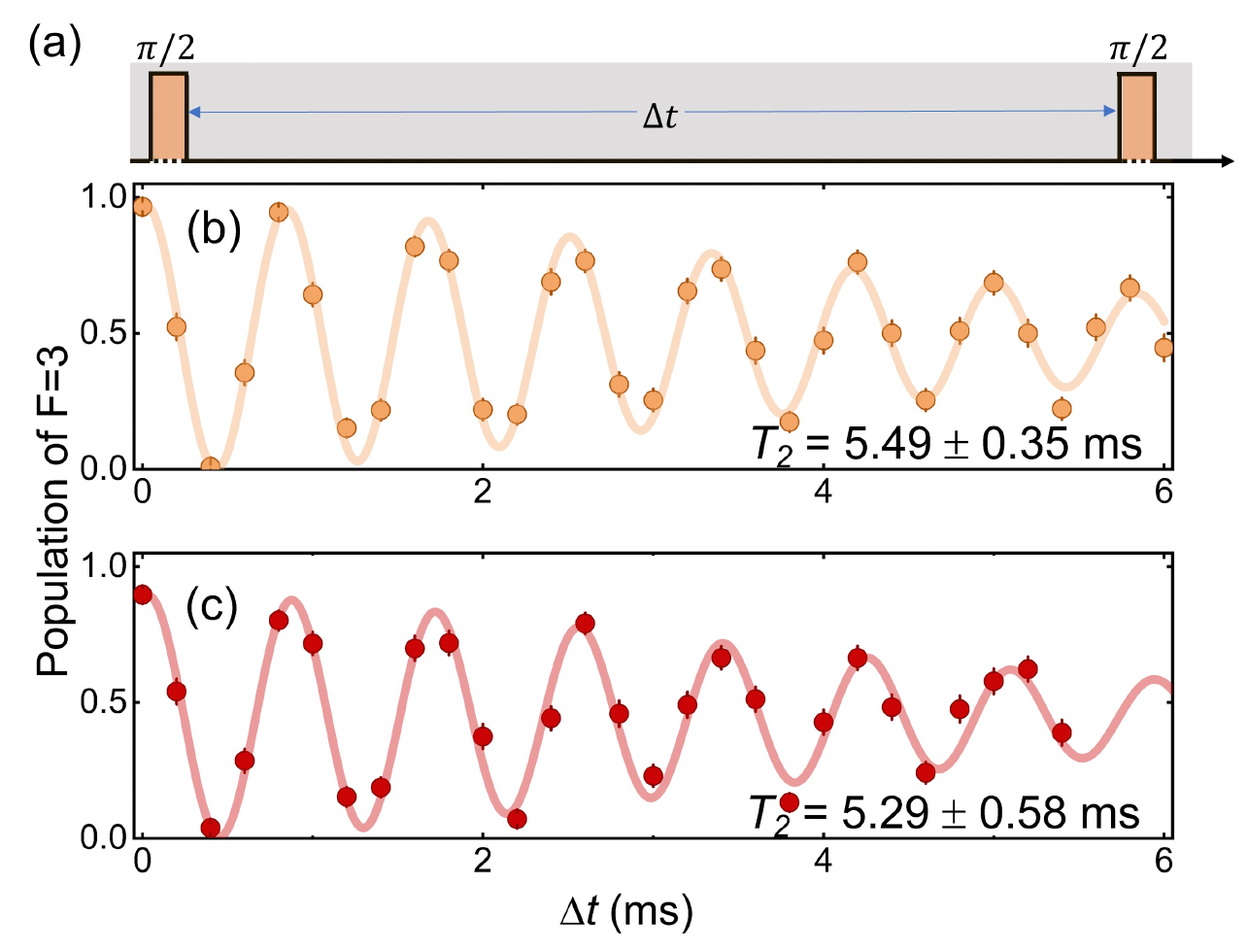}
\caption{Microwave pulse sequence for the Ramsey interferometer and Ramsey fringes. (a) Time sequence of the Ramsey interferometer. (b) A Ramsey fringe when the 1052-nm laser is free running. (c) Ramsey fringe when 40-dB intensity noise is added to the 1052-nm trap laser.}
\label{SM1}
\end{figure}

The Ramsey fringe with the 1052-nm trap laser free running and with 40-dB intensity noise added are taken, and the results are shown in  Fig. \ref{SM1}.
The data fitting by the formula in \cite{Kuhr2005} gives the $1/e$ coherence times of $5.49\pm 0.35$ and $5.29\pm 0.58$ ms, respectively. The temperature of the trapped atom can be inferred by \cite{Kuhr2005}
\begin{equation}
T_2^*=0.97\frac{2\hbar}{\eta k_BT}, \label{T2*}
 \tag{S6}
\end{equation}
and are 17.6 and 18.3 $\mu$K, respectively. The temperature is slightly higher than that (10 $\mu$K) measured by the release and recapture measurements. We therefore estimate the PJR by using the average temperature with  $T\approx 14 $ $\mu$K. The noise levels in
Table \ref{tab1} is also used for the estimation. The estimated PJRs for the free-runing and 40-dB-noise-added traps are 0.5 and 6.5 s$^{-1}$, respectively.

\begin{table}
\caption{\label{tab1} The trap frequencies in the $r$ ($x$ and $y$) and $z$ directions and the corresponding relative intensity noise (RIN) levels used for estimation of the PJR.}
\begin{ruledtabular}
\begin{tabular}{ccccc}
Condition & $f_r$ (kHz) & $S_k(2 f_r) $ (dBc)& $f_z$ (kHz) & $S_k(2 f_z)$ (dBc)\\
\hline
\begin{tabular}{c}
Free running\\
40-dB noise
\end{tabular}
& 30.3 &
\begin{tabular}{c}
$-146$\\
$-104$
\end{tabular}
& 2.7 &
\begin{tabular}{c}
$-110.5$\\
$-103.5$
\end{tabular}

\end{tabular}
\end{ruledtabular}
\end{table}

\begin{figure}[t]
\centering
\includegraphics[width=0.5\columnwidth]{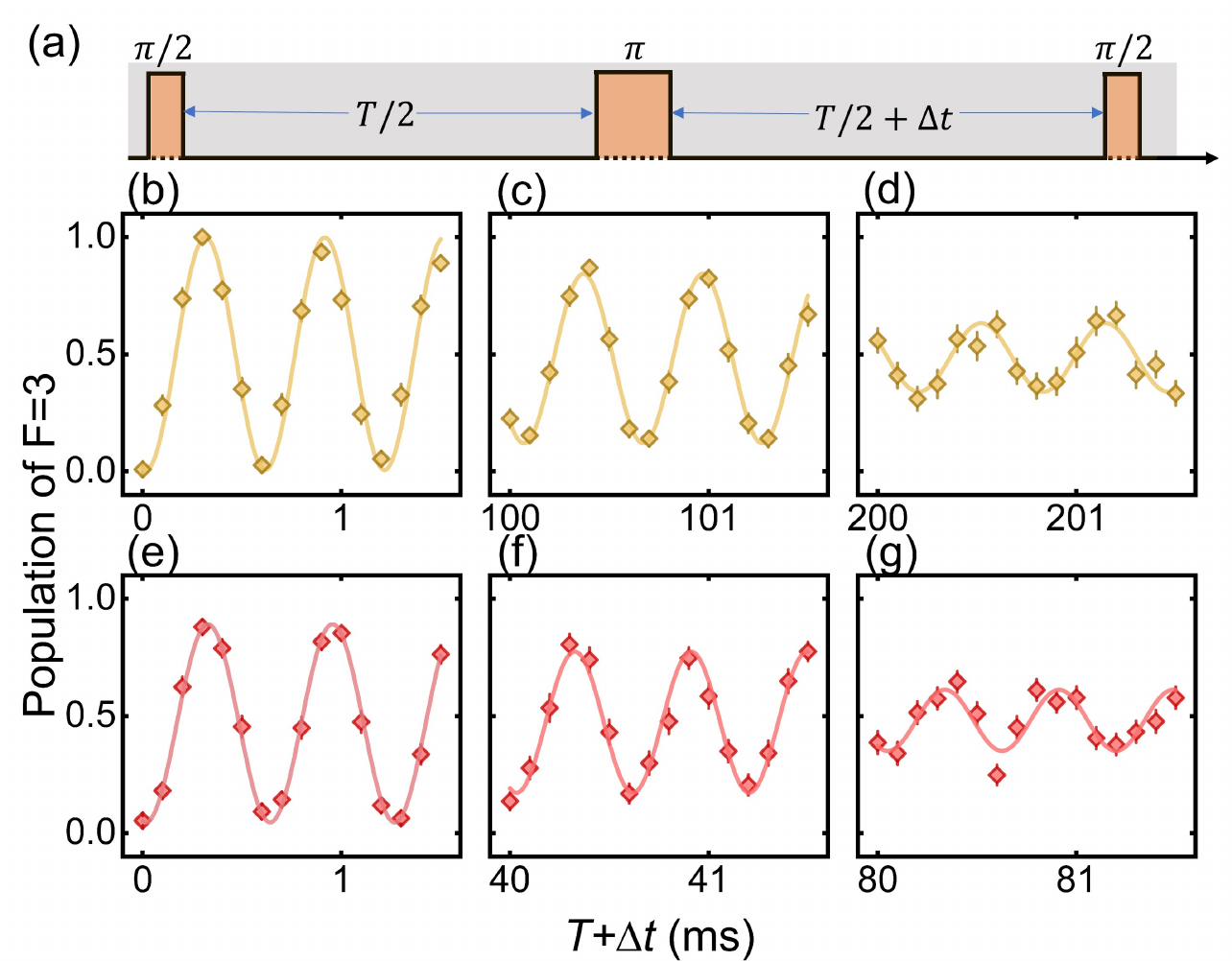}
\caption{Microwave pulse sequence of the spin-echo interferometer and the interfering fringes. (a) The sequence of the spin-echo interferometer. (b)-(d) The spin-echo fringes when the ODT laser is running freely. (e)-(g) The spin-echo fringes when 40-dB intensity noise is added to the ODT laser.}
\label{SM2}
\end{figure}

The data shown in Fig. 2 of the main text come from the fitting of the spin-echo fringe at a series time delay $T$ by sine functions. The fitted amplitude is normalized to the one with $T=0$, which represents the coherence value at time $T$. The typical spin-echo fringes are shown in Fig. \ref{SM2}. Figure \ref{SM2}(a) shows the time sequence for the spin echo interferometer. Figure \ref{SM2}(b)-(d) show the spin-echo fringes with laser free running and with time delays of $T=0$, 100, and 200 ms, respectively. Figure \ref{SM2}(e)-(g) show the spin-echo fringe with 40-dB intensity noise added to the ODT laser and with time delays of $T=0$, 40, and 80 ms, respectively.

\section{Experimental details of the single Cs atom in the 780-nm BBT}

\begin{figure}[t]
\centering
\includegraphics[width=0.75 \columnwidth]{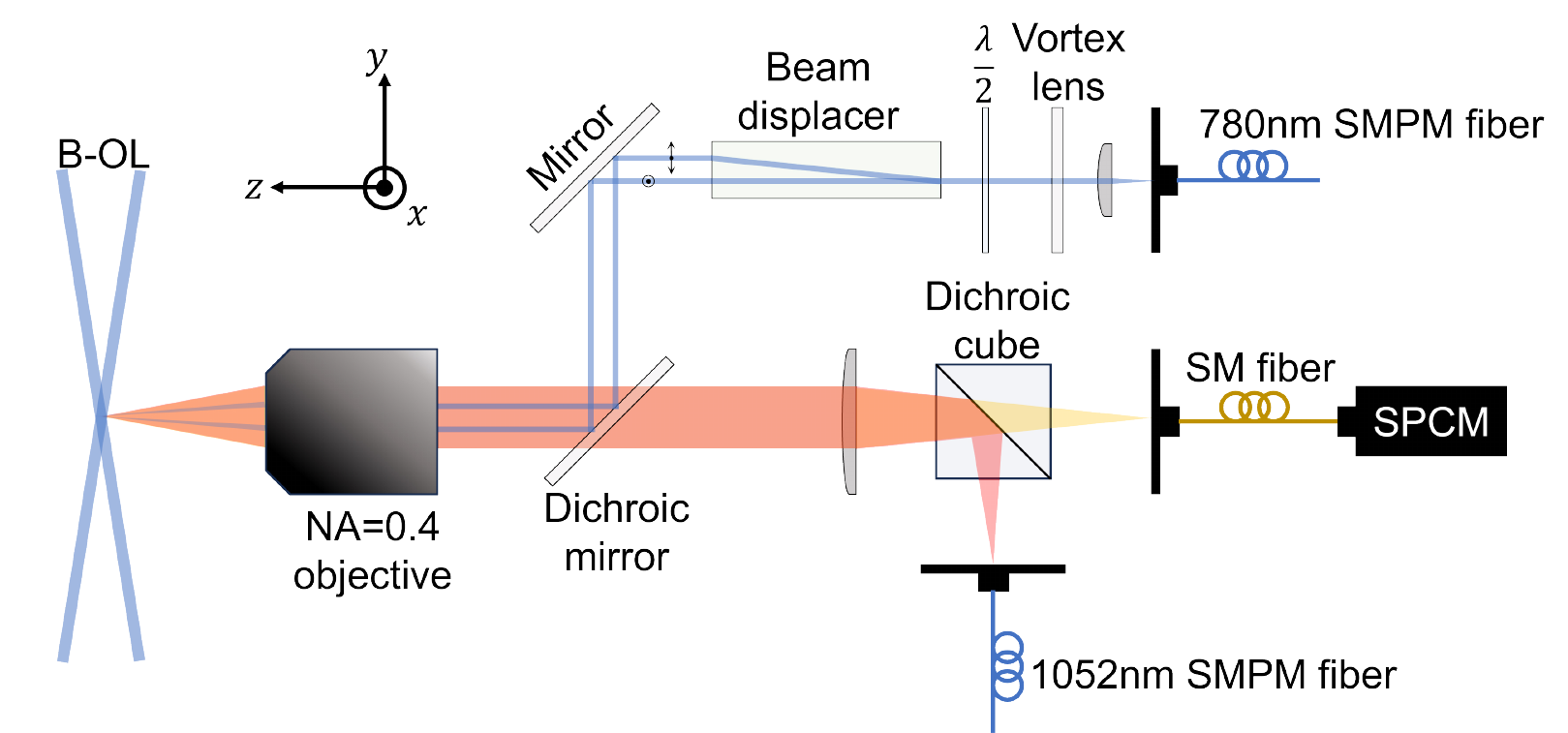}
\caption{The experimental sketch for single atom manipulation in the blue-detuned bottle beam trap (BBT). SMPM fiber: single-mode polarization-maintaining fiber. SPCM: single photon counting module. }
\label{BBT}
\end{figure}

A long coherence of a single Cs atom is achieved in a blue-detuned bottle beam trap (BBT), which is built by crossing two 780-nm Laguerre-Gaussian beams \cite{Li2012}. Figure \ref{BBT} shows a sketch of the experimental setup. First, a 780-nm beam in LG$_{00}$ mode with waist radius $w_0= 2.48$ mm is converted to LG$_{01}$ mode by a vortex lens (HOLO/OR VL-209-M-Y-A). Then, the beam is separated into two parallel beams with equal power and orthogonal polarization by a beam displacer. The distance between the two beams was 4 mm. Finally, the two beams are focused by an NA=0.4 objective, and a BBT is obtained at the focal point. By using a total power of 9 mW, we can construct a trap with a minimum barrier height of $k_B \times 50 $ $\mu$K. The trap frequencies are $(\omega_x,\omega_y,\omega_z)=2\pi \times(5.65,8.30,0.435)$ kHz. 

To efficiently prepare a single atom to three-dimensional (3D) motional ground states, Raman sideband cooling is applied in a combined trap composed of a red-detuned ODT (R-ODT) and a blue-detuned optical lattices (B-OL), which vastly increase the constraint in all three directions. The trap centers of the BBT and combined traps overlap spatially. The trap frequencies of the combined traps are $(\omega'_x,\omega'_y,\omega'_z)=2\pi \times(69.7,59.5,32.3)$ kHz, and the corresponding Lamb-Dicke parameters are $(\eta_x,\eta_y,\eta_z)=(0.172,0.186,0.253)$. After 50 Raman sideband cooling cycles, 82\% of the Cs atoms populate their three-dimensional ground states \cite{Tian2023_RSC}. 

The time sequence of the experiment is shown in main text Fig. 3(d). A single Cs atom is first loaded by the R-ODT from a cold atomic ensemble, which is prepared by a MOT. Then, the B-OL is switched on to compress the confinement of the loaded atom in the axial direction. A resolved Raman sideband cooling (RSC) phase is followed to prepare the atom in its 3D ZPS, and the residual phonon numbers are $(\bar{n}_x,\bar{n}_{y},\bar{n}_z) \simeq (0.07\pm0.07,0.04\pm0.04,0.08\pm0.06)$ \cite{Tian2023_RSC}. Next, the single atom in 3D ZPS is transferred to the BBT by adiabatically switching off the combined ODT. At this time, the atom stays in state $|6S_{1/2} F=4, m_F=4 \rangle$. It is then transferred to $|6S_{1/2} F=4, m_F=0 \rangle$ by four microwave $\pi$-pulse via the intermediate states $|6S_{1/2} F=3, m_F=3 \rangle$, $|6S_{1/2} F=4, m_F=2 \rangle$, and $|6S_{1/2} F=3, m_F=1 \rangle$, and the total transfer efficiency is approximately 96\%. Spin-echo between the Cs clock states ($|6S_{1/2} F=3, m_F=0 \rangle$ and $|6S_{1/2} F=4, m_F=0 \rangle$) is performed to evaluate the decay of the coherence.  The atom state is finally detected.

\begin{figure}[t]
\centering
\includegraphics[width=0.5\columnwidth]{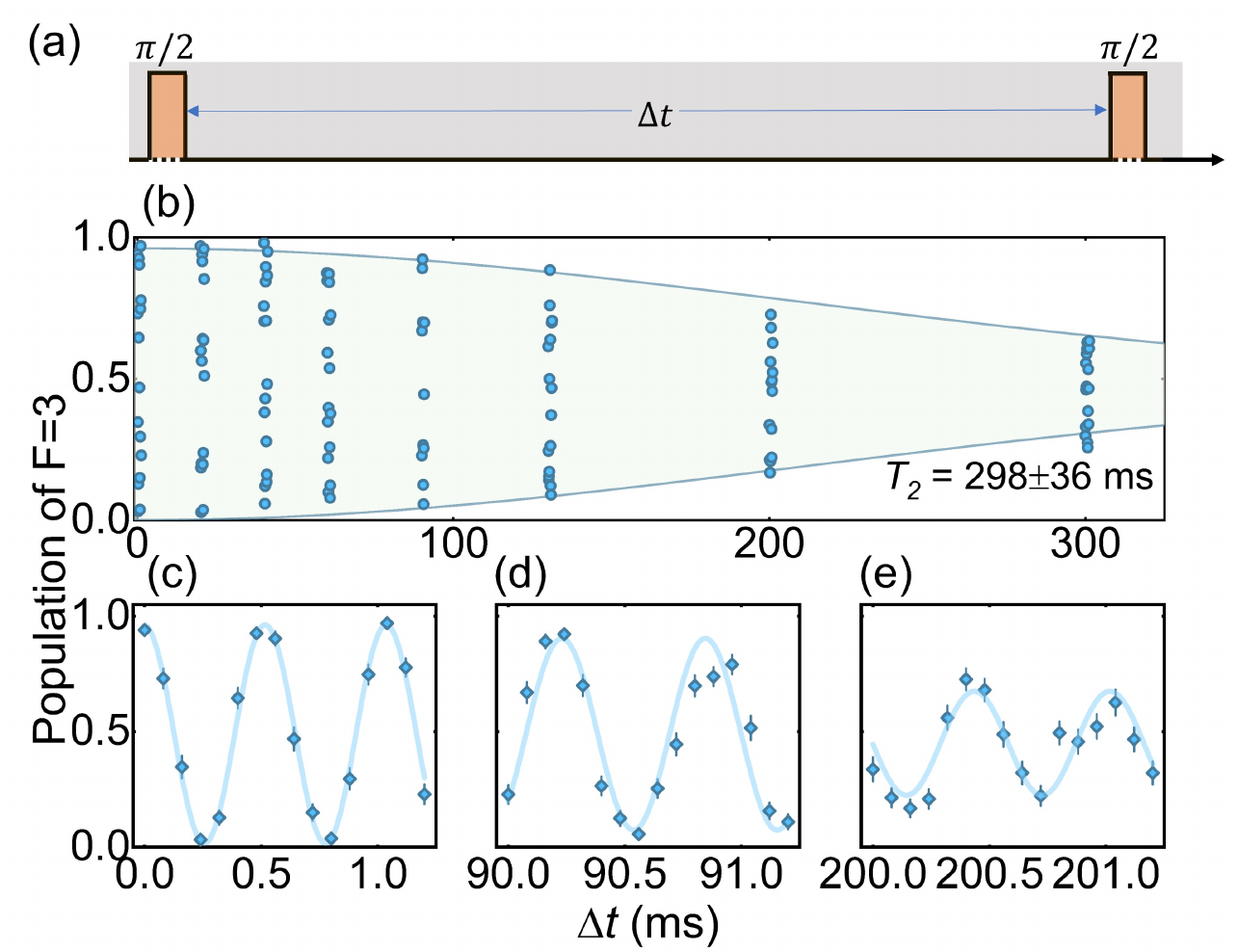}
\caption{Pulse sequence for the Ramsey interferometer and the interfering fringe of single Cs atom in the BBT. (a) Pulse sequence of Ramsey interferometer. (b) The measured Ramsey fringe. (c)-(e) Enlarged view of the data with time delays of $T=0$, 90, and 200 ms.}
\label{SM4}
\end{figure}

\begin{figure}[t]
\centering
\includegraphics[width=0.5\columnwidth]{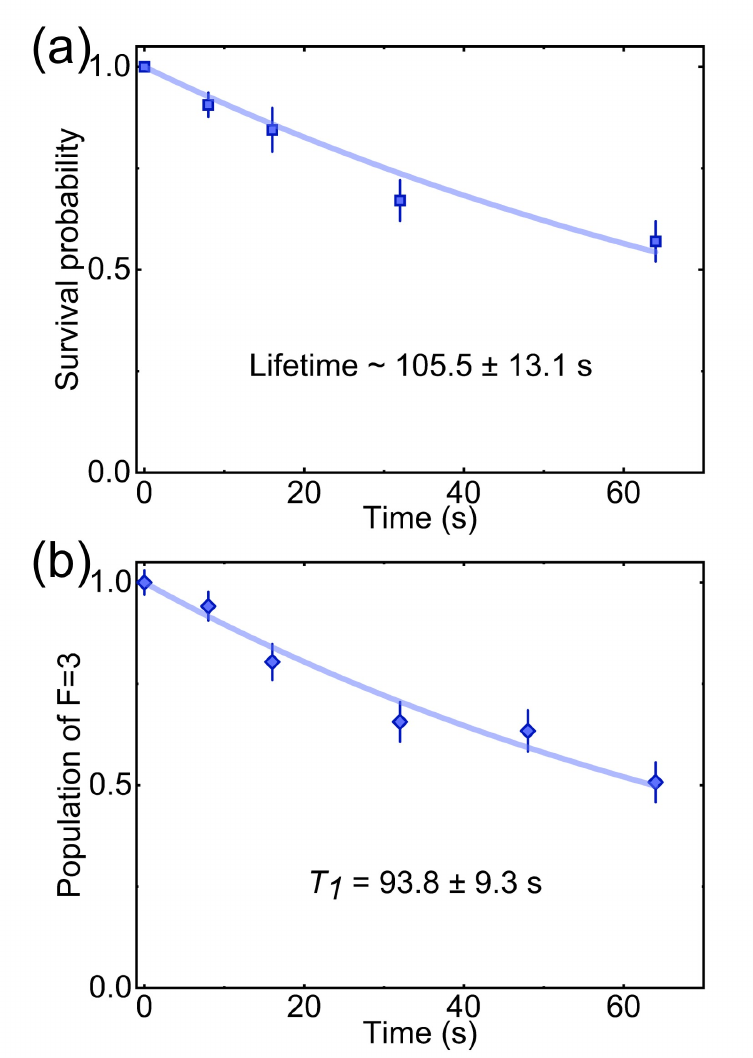}
\caption{(a) The atom survival probability versus the time. The exponential data fitting gives an atom lifetime of $105.5 \pm 13.1$ s. (b) The population of atoms in state $|6S_{1/2}F=3\rangle$ versus time. The exponential data fitting gives a state lifetime $T_1=93.8 \pm 9.3$ s.}
\label{SM3}
\end{figure}

The lifetime of the trapped Cs atom and the state lifetime $T_1$ in BBT were also measured, and the results are shown in Fig. \ref{SM3}. The survival probability of the atom in the trap decreases exponentially with increasing holding time [Fig. \ref{SM3}(a)]. A lifetime of 105.5$\pm$13.1 s is then obtained by the data fitting. The atom loss is predominated by the collision of the residual gas in a vacuum. An ultrahigh vacuum (on the order of $10^{-9}$ Pa) guarantees a hundred-second lifetime of a single atom. We also prepare single atoms for the $|6S_{1/2}F=3,m_F=0\rangle$ state, and observe the population of state $|6S_{1/2}F=3\rangle$ versus the holding time [Fig. \ref{SM3}(b)]. We finally obtain a state lifetime $T_1=93.8\pm9.3$ s, which is mainly limited by the atom lifetime.

\begin{figure}[t]
\centering
\includegraphics[width=0.5\columnwidth]{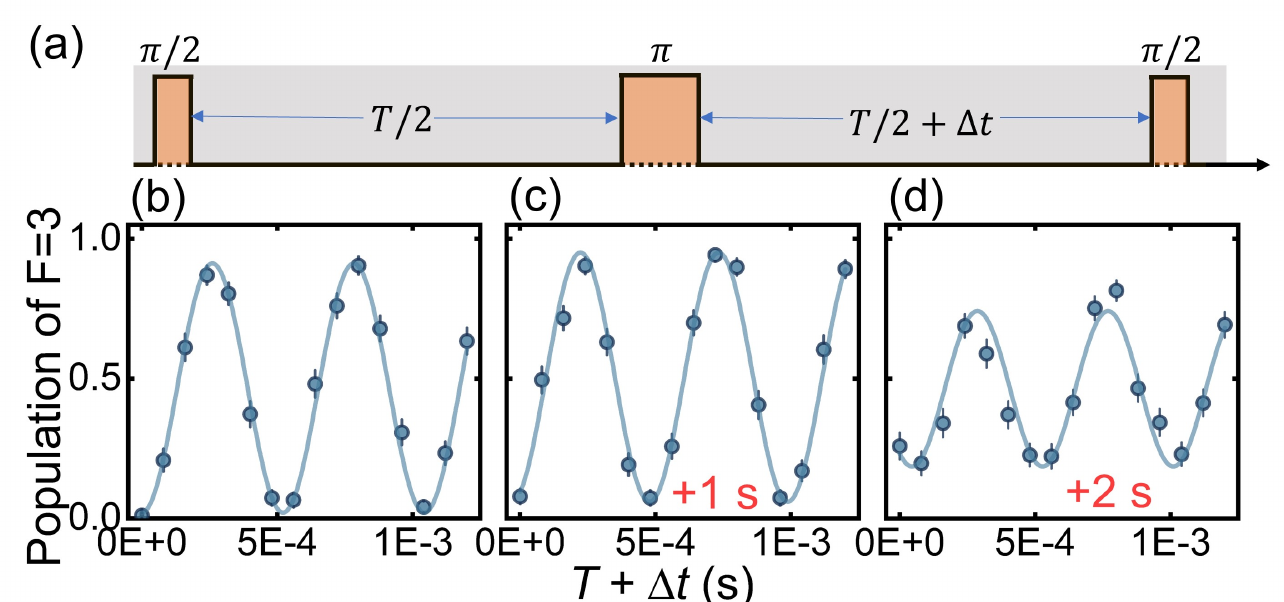}
\caption{(a) Microwave pulse sequence of the spin-echo interferometer and (b-d) the interfering fringe of a single CS in the BBT.}
\label{SM5}
\end{figure}

The figures in Fig. \ref{SM4} demonstrate the Ramsey fringes between states $|6S_{1/2}F=4,m_F=0\rangle$ and $|6S_{1/2}F=3,m_F=0\rangle$. The pulse sequence is similar to that used in R-ODT [Fig. \ref{SM4}(a)]. A coherence time of $298\pm36$ ms from the Ramsey fringe was extracted. The corresponding temperature can be deduced from Eq. (\ref{T2*}) with $T=200$ nK, which agrees with the temperature of the motional ground states in the BBT. The spin-echo fringes are also taken and some of the results are shown in Fig. \ref{SM5}. The fringe amplitudes are extracted by fitting with sine functions. The amplitudes are normalized to that at $T=0$ and the decays with time delay $T$ are summarized in Fig. 4(a) of the main text.

\begin{figure}[t]
\centering
\includegraphics[width=0.5\columnwidth]{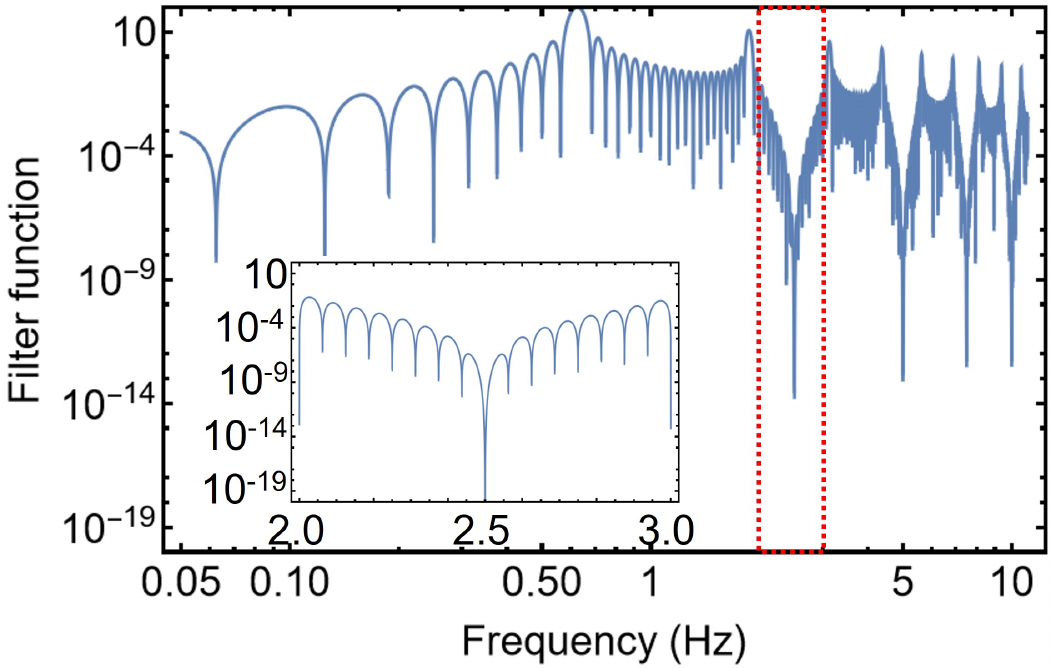}
\caption{Filter function of the CPMG pulse sequence. The insert shows the details of the region within the red dashed box.}
\label{SM6}
\end{figure}

The coherence in the main text is obtained by Carr-Purcell-Meiboom-Gill (CPMG) decoupling sequence. The CPMG sequence eliminates the effects of noise at a particular frequency by periodically reversing the phase of evolution, which acts like a filter \cite{Dynamical_decoupling}. The time sequence of the CPMG used in our experiment is displayed in Fig.3(c) of the main text. In our experiment, a reversing period of 0.8 s was used. The sequences filter out the noise at frequencies $n\times 0.0625$ Hz ($n=1,2,3,\cdots$), especially at the frequencies of $n\times 2.5$ Hz ($n=1,2,3,\cdots$). Figure \ref{SM6} demonstrates the filter function of our CPMG sequence.

\section{Photon-scattering induced decoherence}

Previously, it was assumed that the coherence between two electronic ground states of an atom can be preserved via Rayleigh scattering. Here we show that even Rayleigh scattering can destroy coherence. The two ground states are denoted by $|a\rangle$ and $|b\rangle$, and $|b\rangle$ is coupled to an excited state $|e\rangle$ by the trapping light field with a Rabi frequency $\Omega$ and one-photon frequency detuning $\Delta$. The excited state $|e\rangle$ decays only to $|b\rangle$ with a photon decay rate $\Gamma$. The Hamiltonian of the system is 
\begin{equation}
H=\frac{\Omega}{2} \left(|e\rangle \langle b| +|b\rangle \langle e|  \right) + \left( \Delta + i \frac{\Gamma}{2} \right) |e\rangle \langle e|. \label{HScattering}
 \tag{S7}
\end{equation}
We then have the Heisenberg equations: 
\begin{subequations}
\begin{align} 
\frac{\text{d} }{\text{d} t} |a\rangle \langle b| &=- i \frac{\Omega}{2} |a \rangle \langle e| , \tag{S8a} \label{HE1} \\
\frac{\text{d} }{\text{d} t} |a\rangle \langle e| &=- i \frac{\Omega}{2} |a\rangle \langle b| -\left( \Delta + i \frac{\Gamma}{2} \right) |a\rangle \langle e|. \tag{S8b} \label{HE2} 
\end{align}
\end{subequations}
Assuming $\frac{\text{d} }{\text{d} t} |a\rangle \langle e| =0$, we then have 
\begin{equation}
\frac{\text{d} }{\text{d} t} |a\rangle \langle b| =-\frac{ \frac{\Omega}{2} \left( \Delta-i  \frac{\Gamma}{2} \right) } {\Delta^2 +\left(\frac{\Gamma}{2}\right)^2} |a\rangle \langle b|. \label{DynCoher}
 \tag{S9}
\end{equation}
When $\Delta \gg \Gamma$, the coherence at any time $t$ is then
\begin{equation}
|a\rangle \langle b| _{t}=\exp{\left[ \left( i \Delta_\text{LS}  - \frac{1}{2} R_\text{s} \right) t \right]}, \label{DynCoher}
 \tag{S10}
\end{equation}
where we set the coherence to 1 at time $t=0$. $\Delta_\text{LS}=\frac{ \Omega^2}{4 \Delta}$ is the light shift of $|b\rangle$, whose fluctuation is the variance of the DLS discussed in the main text. $R_\text{s}=\frac{ \Omega^2}{4 \Delta^2} \Gamma$ is the photon scattering rate, which limits the coherence time. We then have a scattering-limited coherence time 
\begin{equation}
T_2^\text{(s)}=2/R_\text{s}. \label{T2s}
 \tag{S11}
\end{equation}
Here we see that the $T_2$ time is limited by the overall scattering rate. 

\end{document}